\documentclass[sigconf]{acmart}

\let\oldhref\href
\renewcommand{\href}[2]{\oldhref{#1}{\hbox{#2}}}

\usepackage{balance}
\usepackage{soul}

\usepackage{graphicx}
\usepackage{tabularx}
\usepackage{booktabs}
\usepackage{url}

\usepackage[center]{subfigure}
\usepackage{comment}
\usepackage{enumerate}
\usepackage{xspace}

\usepackage{multirow}

\usepackage{amsthm}
\usepackage{amsfonts}
\usepackage{mathtools}
\usepackage[utf8]{inputenc}

\usepackage{color}

\usepackage{enumitem}
\setlist[itemize]{noitemsep, topsep=0pt, leftmargin=10pt}

\usepackage{colortbl}
\definecolor{Gray}{gray}{0.65}
\definecolor{LightGray}{gray}{0.9}

\newcommand{\todosd}[1]{\textcolor{Green}{\small{S: #1}}} 

\newcommand{\lfig}[1]{\label{fig:#1}}

\newcommand{\rfig}[1]{Fig.~\ref{fig:#1}}

\newcommand{\fakeparagraph}[1]{\vskip 0pt\noindent\textbf{#1 }}

\setcopyright{acmlicensed}

\acmDOI{X}

\acmISBN{}

\copyrightyear{2017} 
\acmYear{2017} 
\setcopyright{acmlicensed}
\acmConference{CCSW'17}{November 3, 2017}{Dallas, TX, USA}\acmPrice{15.00}\acmDOI{10.1145/3140649.3140656}
\acmISBN{978-1-4503-5204-8/17/11}

\begin{document}
\title{Towards Blockchain-based Auditable Storage\\ and Sharing of IoT Data}

\author{Hossein Shafagh}
\affiliation{\institution{Department of Computer Science\\ ETH Zurich, Switzerland}}
\email{shafagh@inf.ethz.ch}

\author{Lukas Burkhalter}
\affiliation{\institution{Department of Computer Science\\ ETH Zurich, Switzerland}}
\email{burkhalter@inf.ethz.ch}

\author{Anwar Hithnawi}
\affiliation{\institution{Department of Computer Science\\ ETH Zurich, Switzerland}}
\email{hithnawi@inf.ethz.ch}

\author{Simon Duquennoy}
\affiliation{\institution{RISE SICS, Sweden}}
\email{simon.duquennoy@ri.se}


\begin{abstract}
Today the cloud plays a central role in storing, processing, and distributing data.
Despite contributing to the rapid development of IoT applications, the current IoT cloud-centric architecture has led into a myriad of isolated data silos
that hinders the full potential of holistic data-driven analytics within the IoT.
In this paper, we present a blockchain-based design for the IoT that brings a distributed access control and data management.
We depart from the current trust model that delegates access control of our data to a centralized trusted authority and instead empower the users with data ownership.
Our design is tailored for IoT data streams and enables secure data sharing.
We enable a secure and resilient access control management, by utilizing the blockchain as an auditable and distributed access control layer to the storage layer.
We facilitate the storage of time-series IoT data at the edge of the network via a locality-aware decentralized storage system that is managed with the blockchain technology.
Our system is agnostic of the physical storage nodes and supports as well utilization of cloud storage resources as storage nodes.
\end{abstract}

\maketitle

\section{Introduction}
With the emergence of networked embedded devices dubbed as the IoT,
we are witnessing an ever increasing number of innovative applications.
The current ecosystem of the IoT consists typically of designated low-power devices equipped with sensors collecting data.
This data is then stored via special-purpose apps (i.e., application-layer gateways) in a third-party cloud storage for further processing.

This stove-piped architecture~\cite{Zachariah} has resulted into isolated data silos,
where users have limited control over their data and how it is used.
Users have to trust the cloud and application providers and have no choice but to rely on their promises of security and availability.

 \begin{figure}[t]
	\begin{center}
	\includegraphics[width=1\columnwidth]{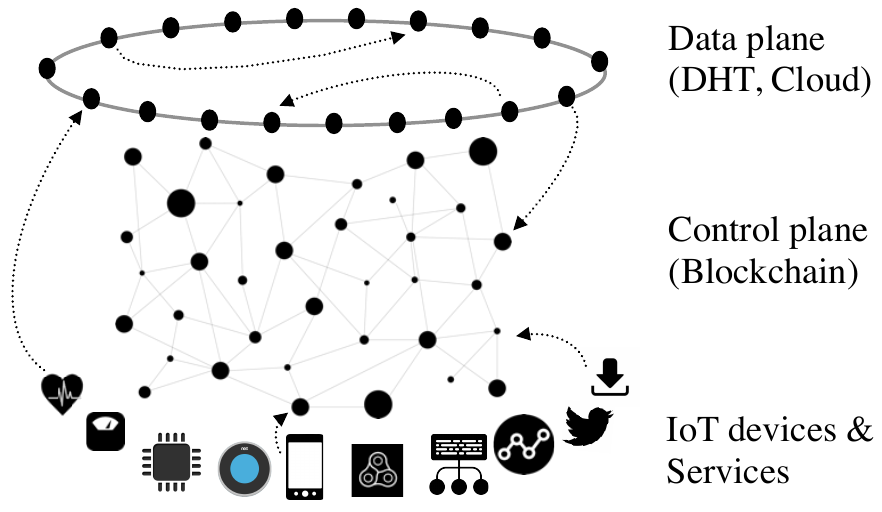}

	\caption{Our blockchain-based, end-to-end encrypted, and decentralized data storage.
	No central trusted authority controls access to users' data.
	}
	
	\lfig{systemoverview}	
	\end{center}
\end{figure}

Today's IoT security efforts mostly focus on securing point-to-point communication and
fall short in addressing security during the life-cycle of data (e.g., auditable access control, secure sharing).
The current cloud-based model in the IoT handles identification, authentication, and connectivity of IoT devices.
Although this model has enabled the bootstrap of the IoT ecosystem, it is not necessarily the most suitable solution for the IoT~\cite{GDP},
{as it neglects the locality of data and mandates centralization through trusted third parties}.

These limitations necessitate a rethinking of the way we currently handle IoT data.
Instead of giving up ownership of our data to various applications,
we enable an independent and resilient data management system that ensures data ownership.
We identify the following requirements for such a system:
\textit{(R1)}~Decentralized, resilient, and auditable access control management (ownership, cryptographically secure sharing);
\textit{(R2)}~Secure data storage (confidentiality, authenticity, integrity);
\textit{(R3)}~IoT compatibility (append-only data streams, with a single writer and several readers).
Conventional cloud-based solutions provide \textit{R2} and when relevant \textit{R3}, but fall short in addressing \textit{R1}.
{Recent decentralized storage startup efforts (Sia~\cite{sia}, Storj~\cite{Storj}, Filecoin~\cite{Filecoin})
show the potential of decentralized  blockchain-based storage with financial incentives.
These efforts, however, are optimized for file storage and fall short in accommodating for time-series data of the IoT (i.e., \textit{R3}).}

We propose a blockchain-based access control management to address \textit{R1}.
This provides us with an independent network that maintains a distributed ledger of access control permissions.
Inspired by recent blockchain-based technologies~\cite{Enigma, Blockstack}, we combine the blockchain with an off-chain storage, for a scalable secure data storage to address \textit{R2}.
Finally, our system is designed from the ground up to support IoT data streams to address \textit{R3}.
Our system accommodates for IoT data streams where
streams are chunked, compressed, and encrypted in the application layer and only authorized services are granted access to the decryption keys.
This requires us to address several challenges for an efficient key distribution and management scheme as well as a secure storage of data stream chunks.
The built-in cryptocurrency feature in the underlying blockchain technology would allow the realization of an autonomous,
self-sustaining decentralized storage ecosystem, where storage nodes are rewarded for providing storage and bandwidth, and more importantly for following the protocol correctly.

The contribution of this paper is the design of a novel blockchain-based auditable data-management system for IoT data.
Our system exhibits 
\textit{(i)}~a cryptographically secure data sharing with frequent key updates,
\textit{(ii)}~the possibility of access revocation,
\textit{(iii)}~an efficient search of compressed chunked data streams, and
\textit{(iv)}~a distributed locality-aware storage layer.

\section{Background}

In this section, we briefly review important aspects of the cloud, the IoT ecosystem, and give a primer on the blockchain technology.

\subsection{Cloud}

Cloud platforms are typically hosted in large-scale data centers that are located at the edge of the Internet backbone~\cite{GDP}.
The consolidation and centralization of data centers, however, yield an increased distance between clients and services.
This results in a high variability in latency and bandwidth.
To address this issue, especially with regards to resource-intensive and interactive applications, decentralized cloud architectures, namely cloudlets, have emerged. 
Cloudlets are small-scale data centers that are located closer to users and can meet low latency and high bandwidth guarantees.
{Our system embraces this locality-aware data storage and processing trend and brings it to its full potential with our decentralized access control layer which ensures ownership and secure sharing of data.}

\subsection{IoT Ecosystem}

Embedded computing devices are increasingly integrated into objects and environments surrounding us.
These devices utilize low-cost sensors for a range of applications. 
The typical system structure of the IoT involves the three tiers of
(i)~low-power IoT devices,
(ii)~a potential gateway that interconnects IoT devices with the Internet, and
(iii)~the backend where IoT data is stored.

IoT devices are typically equipped with resources in the orders of few MHz of CPU, few 10s of KB of RAM, and few 100s of KB of ROM.
Additionally, they can embed low-power hardware crypto accelerators, enabling a new class of secure applications~\cite{talos, Pilatus}, for instance, lightweight clients of a blockchain network.
However, conventional security solutions for the IoT still utilize pre-shared symmetric keys for the secure communication.
This simple approach does not scale for the massive number of IoT devices.
Efforts~\cite{hummen2014delegation} to tailor public-key based secure communication to the IoT remain to find widespread adoption.
{Leveraging the blockchain technology, we enable a decentralized management of identities of IoT devices and enable a transparent device ownership.}

\subsection{Blockchain}

A blockchain is essentially  a distributed ledger that consists of a continuously growing set of records. 
The distributed nature of blockchains implies no single entity controls the ledger (i.e., censorship/suppression resistant),
but rather the participating peers together validate the authenticity of records.
These records are organized in blocks which are linked together using cryptographic hashes, hence the name blockchain.
Blockchain-based technologies~\cite{NBFMG16} incentivize a network of peers to make computations towards consensus in the network.

The most prominent example of a successful blockchain deployment 
is the Bitcoin cryptocurrency (the decentralized peer-to-peer digital currency)~\cite{nakamoto2008bitcoin, BitCoin:SoK}. 
The Bitcoin blockchain maintains all transactions from the initial block, referred to as the genesis block.
A transaction contains the sender, receiver, amount of the transferred Bitcoin currency, and signature of the sender.
For a transaction to be included in the blockchain (i.e., to be considered as valid), it is transmitted to the blockchain network.
The so-called miners take the responsibility to verify new transactions and suggest the next block which includes the verified transactions.
Miners are rewarded with Bitcoins and transaction fees for their computational work.

To prevent a single miner from dominating the blockchain network and hence having the power of manipulating the history of transactions,
the concept of
proof-of-work~\cite{nakamoto2008bitcoin} is employed to reach consensus in the blockchain network. 
A new block includes a set of new transactions, the hash of the previous block,
the miner's address who is suggesting this block, and most importantly 
the answer to a difficult-to-solve mathematical puzzle.
This mathematical puzzle is unique to each block and easy to verify once found.
Once a miner finds such a block, it publishes it such that all nodes and miners can verify its correctness and consider it as the new valid block to build upon.
In case several valid blocks are suggested at the same time, miners randomly select the next block.
Eventually, the network converges towards the longest branch of the blockchain as the main branch.
Solving the puzzle is referred to as the proof-of-work and ensures as well resistance against Sybil attacks.

{Bitcoin and its most prominent contender Ethereum~\cite{ethereum} are permission-less blockchains} where any node can become a miner or just a client.
Permissioned blockchains, such as the hyperledger~\cite{Hyperledger}, allow a designated set of authorized validator nodes (i.e., miners) to participate in the block validation process.
Such blockchains typically use more CPU-friendly consensus protocols, such as the Practical Byzantine Fault Tolerance protocol~\cite{PBFT},
since the set of validator nodes is known.
Hence, permissioned blockchains can handle a higher transaction throughput (7 vs. $10^4$ transactions per second).
However, permissioned blockchains require a trusted central party to initially authorize the blockchain validators.\linebreak
Moreover, due to the high communication overhead, i.e., $O(n^2)$, only deployments of a few tens of validators are practical.

\section{System Design}
In a nutshell, we decouple the control and data plane of our IoT distributed storage system (see \rfig{systemdesign}).
We realize the access control layer using a public blockchain, to satisfy \textit{R1}.
Bitcoin is our current candidate for the blockchain layer {in our reference implementation} due to its strong security, reliability, and current dominance.
However, other cryptocurrencies~\cite{sasson2014zerocash} can be employed seamlessly.
This is possible, because our system's logic resides in a virtualchain~\cite{Virtualchain, Blockstack} and outside the blockchain.
Virtualchain allows the introduction of new functionality to production blockchains, without requiring any changes in the underlying blockchain.

The data plane consists of a routing layer and the 
secure storage layer to satisfy \textit{R2}.
The storage layer is composed of either an on-premises storage, the cloud, or a distributed peer-to-peer network.
Data is encrypted end-to-end at the client-side.
Hence, the storage nodes have no insights about the hosted data at their side.\linebreak
The data in our system is structured in streams, to accommodate for IoT-specific needs (statisfies \textit{R3}).
In concrete terms, ownership and sharing permissions are per stream, and streams are chunked and encrypted before storage.

In the following, we detail our design for the control plane and our data plane features.

 \begin{figure}[t]
	\begin{center}
	\includegraphics[width=1\columnwidth]{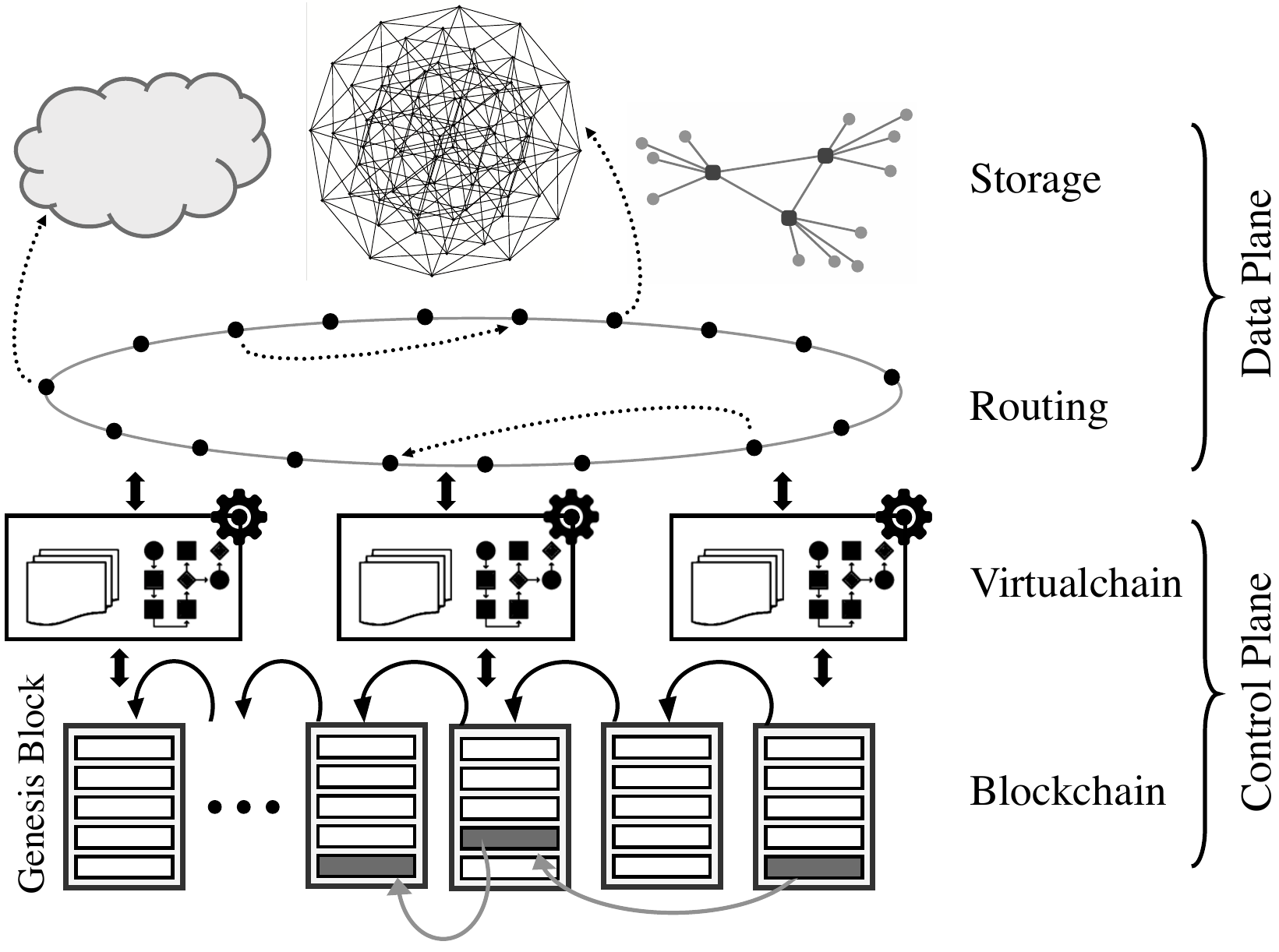}

	\caption{Overview of our layered design. 
	Transactions in the blockchain can contain access permissions (gray).
	}
	
	\lfig{systemdesign}	
	\end{center}
\end{figure}

\subsection{Control Plane}

In our system, the control plane is logically separated and agnostic of the data plane.

\fakeparagraph{Blockchain.}
We employ a publicly verifiable blockchain to create an accountable distributed system and
bootstrap trust in an untrusted network, without a central trust entity.
In our system, transactions consist of ownership of data streams and corresponding access permissions.
Our access control transactions, similar to default transactions of the underlying cryptocurrency, remain publicly auditable
(see~\rfig{systemdesign}).
{To preserve the privacy of access permissions, we can rely on stealth addresses~\cite{stealth}.}

\fakeparagraph{Access Control.}
We use the blockchain to store access permissions securely.
Access rights are granted per data stream and
the data owner can revoke the sharing of a data stream.
Initially, the data owner issues a transaction including the stream identifier (i.e., hash digest).
To share the data stream with a service, the data owner issues a new transaction which holds 
\textit{(i)}~the stream identifier and
\textit{(ii)}~the public key address of the service.

For any request to retrieve data, the storage node first checks the blockchain for access rights.
Note that a malicious storage node could hand out data without permission.
However, the impact of this action is limited since
\textit{(i)}~data is encrypted,
\textit{(ii)}~in the case of DHT, each node holds a small random fraction of a data stream.
{Moreover, economic incentives (i.e., collateral and reward) should encourage storage nodes to follow the protocol correctly.}

\fakeparagraph{Key Management.}
We enable a low-cost key renewal with key regression~\cite{fu:keyregression}.
In key regression, given key $K_t$ in current time $t$ one can compute all keys until the initial key $K_0$.
This allows us to update the encryption keys frequently, and only share the latest $K_t$ with the sharing services.
However, given $n$ services, this requires a communication overhead in the order of $O(n)$:
at each key update, the key must be shared $n$ times (after encrypting it with the corresponding service's public key).

We propose to employ a re-encryption-based technique to bring the communication overhead to $O(1)$.
Given a re-encryption token $T_{a\rightarrow b}$, one can re-encrypt a ciphertext under Alice's public key $PK_a$ to a ciphertext under Bob's public key $PK_b$, without access to the plaintext~\cite{BBS:98, Ateniese:NDSS}.
To share $K_t$ with all services, Alice encrypts $K_t$ with a one-time public key pair $(PK_a, SK_a)$.
For all services $S_i$, she issues a re-encryption token $T_{SK_a\rightarrow PK_{S_i}}$ based on their public keys $PK_{S_i}$
(this step takes place while issuing the sharing transaction).
Each service $S_i$ can then re-encrypt ENC$_{PK_a}$($K_t$) to ENC$_{PK_{S_i}}$($K_t$), and use their respective secret keys $(SK_{S_i})$ to access $K_t$.
After this point, Alice only needs to update ENC$_{PK_a}$($K_{t+1}$) for the services to preserve their access to the latest key $K_{t+1}$.

\fakeparagraph{Revocation.}
To revoke access to a data stream, the data owner updates the encryption key to $K_{t+1}$.
She then updates the encrypted shared key for authorized readers, however, with a new one-time public key pair: ENC$_{PK_{a'}}$($K_{t+1}$).
Revocation causes a communication overhead of $O(n)$,
since Alice needs to update all valid re-encryption tokens $T_{SK_{a'}\rightarrow PK_{S_i}}$, excluding the revoked service.
This prevents the revoked user to decrypt any future data.

As an additional protection, and for auditing purposes, the user issues a new blockchain transaction overriding previous permissions.
Storage nodes will even decline sharing older data, that the user once had access to.
The impact of a potential dishonest node leaking old encrypted data chunks is low, as old data might have been cached at the user anyway.
New data, however, is protected cryptographically after a key update.

 \begin{figure}[t]
	\begin{center}
	\includegraphics[width=1\columnwidth]{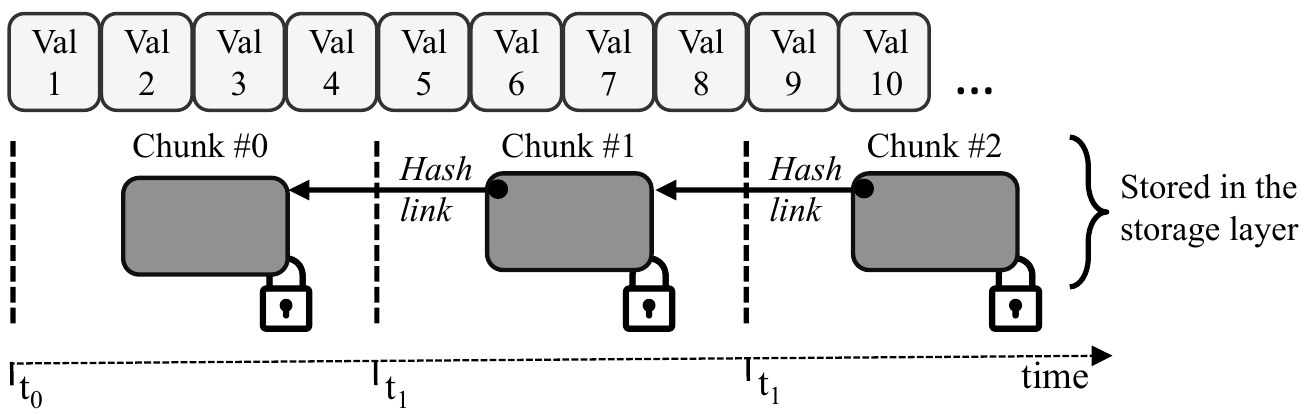}
	\caption{Data streams are chunked at pre-defined lengths, compressed, and encrypted.
	To lookup a record, a local index maps the key of the record to the chunk key.}
	
	\lfig{datachunking}	
	\end{center}
\end{figure}

\subsection{Data Plane}
In order to address \textit{R3}, we consider IoT data types of stream character
where data records are generated continuously, as depicted in \rfig{datachunking}.
Current distributed storage approaches~\cite{ipfs, Storj, Filecoin} primarily target archiving data and are not suitable for IoT data.\linebreak
{Moreover, they either consider data to be public (e.g., IPFS\cite{ipfs}) or store encrypted data without a secure sharing feature (e.g., Storj~\cite{Storj} and Filecoin~\cite{Filecoin}).}

{To store time-series data in our system, we store data chunks which compose several consecutive data records, instead of storing individual data records.}
To this end, we split a data stream into data chunks {which are cryptographically chained together (i.e., each chunk holds a hash pointer to the previous chunk).}
Although chunking data prevents random access at the record level,
there is a positive gain
on the performance of data retrieval since in time-series data most queries require data that is co-located in time (e.g., all records of one day)~\cite{Bolt}.

Note that the data itself is stored off-chain and only its identifier (i.e., hash pointer) is included in the blockchain, ensuring data immutability.
Since adding an identifier for each chunk to the blockchain would not scale, our system adds only data chunks at given intervals into the blockchain.
Due to the fact that all chunks are cryptographically chained together, all chunks that are between two intervals without their identifier in the blockchain become immutable too.
The interval-time corresponds to the maximum time chunks need to become immutable. 
It is tunable and defined by the application logic.

\fakeparagraph{Encryption.}
Each data chunk is encrypted at the source with an efficient symmetric cipher.
We rely on AES-GCM, as an authenticated encryption scheme.
Our chunks have a plaintext field containing the key value of the chunk and the encrypted compressed data records. 
With authenticated encryption, both fields are integrity protected and authenticated.
Services with access to the symmetric data stream key $K_t$ can verify the integrity of the chunk and perform an authenticated decryption.

To ensure data ownership, for instance towards the storage layer, each chunk is in addition signed.
This allows parties without access to the stream key to still be able to verify the owner of the data stream, albeit at a higher computation cost.
Each chunk contains the unencrypted stream identifier linking it to the corresponding access control transactions.

\fakeparagraph{Compression.}
IoT data is highly compressible due to high correlation in time.
Hence, we compress data chunks before encryption.
This reduces bandwidth and storage requirements significantly.
Our initial results of IoT data compression show that even with small chunk sizes, we can reach compression ratios close to the optimum (i.e., compression of the entire data set).
As depicted in \rfig{compression}, compressing the data record of one year by Fitbit\footnote{
Fitbit heart rate $\&$ fitness wirstband: \url{www.fitbit.com/charge2}
}
results into a compression ratio of 9.75 (11.45 for Ava\footnote{
Ava: ovulation tracking bracelet: \url{www.avawomen.com}
}
).
Already with a chunk size of 2048 (corresponding to one day worth of data records for Ava),
we can reach a ratio of 11.08 for encrypted and compressed chunks.

\fakeparagraph{Search.}
In the storage layer, we store key-value pairs.
In our case, the value is the current data chunk of a data stream,
where the key (i.e., a 256-bit identifier) is the cryptographic hash of
the tuple: $<$stream-ID, owner-ID, timestamp-hash$>$.
The IDs are unique bit strings (i.e., hash digests).

To enable an efficient search and query of any record in the data stream, we use a simple technique based on the 
the timestamp $t_0$ of the first chunk and the length of the chunks 
$\Delta$.
To look-up a record with timestamp $t_i$, we compute the timestamp of the chunk holding it.
For instance, the look-up of value 6 in \rfig{datachunking} is mapped to the key of chunk $\#$1.

 \begin{figure}[t]
	\begin{center}
	\includegraphics[width=1\columnwidth]{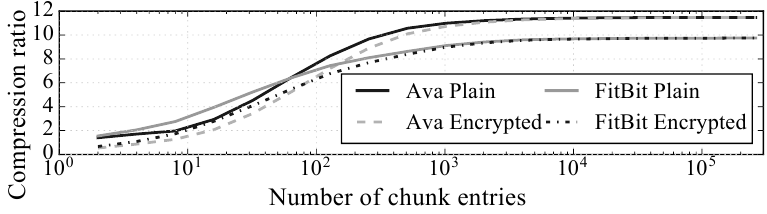}

	\caption{Compression ratio of our chunking with the authors Fitbit data and anonymized Ava
	 data of one year.
	}
	
	\lfig{compression}	
	\end{center}
\end{figure}

\fakeparagraph{Data Storage.}
We advocate a distributed data storage layer, however our design is agnostic of the storage layer.
Hence, on-premise storage and storage on cloud services are compatible with our system.

The IoT gateway serves as an intermediate storage node at the front of the storage layer.
The gateway can push the chunks in a FIFO principle into the storage layer to maintain a reasonable local storage size.
Hence, the IoT gateway serves as a cache of recent data items for 
on the gateway hosted apps and could as well be queried by services in network proximity to the IoT gateway.
Current IoT apps, first push all new data to the cloud and then fetch them again for presentation (e.g., FitBit), introducing an unnecessary latency.

For the distributed storage, we rely on a P2P overlay routing technique and a \textit{Distributed Hash Table} (DHT) as our general-purpose private key-value data store interface.
The DHT serves as a scalable, self-managing storage with high availability
(i.e., robust against targeted communication outages or malicious attacks in case of central servers).
The DHT in general enforces replicated and randomized storage across a 160-bit address space.
However, we rely on the sloppy hashing technique~\cite{coral, freedman2003sloppy} to augment our instantiation of DHT~\cite{Kademlia, SKademlia} with locality.
To this end, chunks are stored/replicated on nodes that are closer to the services or the data owner.

How to financially incentivize participation in a distributed storage is out of the scope of this paper.
Several researchers and a few start-ups propose systems where users can make money with ``renting out" their local storage space~\cite{Storj, Filecoin, maidsafe}.
For instance, proof-of-retrievability~\cite{Storj} can be used as a mechanism to reward users who store more files for a longer time.
Moreover, the reward encourages storage nodes to follow the protocol correctly, for example with regards to enforcing the access permissions.

\subsection{Privacy \& Security Analysis}
For an adversary to alter access permissions in the blockchain it requires forging a digital signature or gaining control over the majority of the compute power in the blockchain network.
The former is prevented with the security of signatures and the latter with
the consensus protocol in the blockchain (i.e., proof-of-work) and its decentralized nature.
Moreover, an adversary is not capable of learning sensitive information from the public blockchain,
since only pseudo-identities and stream identifiers are stored there.
Data chunks are encrypted, integrity protected, and authenticated.
An adversary with access to encryption keys cannot alter stored chunks, as it requires gaining access to the public-private key pair of the data owner.
{Even in case the owner's key is leaked, chunks cannot be modified due to the blockchain immutability (except for chunks in the current interval).
Rational storage nodes follow the protocol correctly due to financial incentives (i.e., interplay of reward and collateral)}

\subsection{Primary Evaluation}
Initial evaluation results from our reference implementation in the bitcoin blockchain (i.e., bitcoin testnet) show reasonable overhead.
For instance, augmenting Amazon's S3 storage with our system's access control results only in a slowdown of 10\% in request throughput.
In comparison, a distributed storage with more than 1000 nodes experiences a factor of 2 slowdown, dominated by the routing.
This slowdown is the worst-case scenario with no locality-awareness in storage of data.
Store and get procedures for individual chunks require 150~ms, assuming no local caching.
We are currently working on thoroughly evaluating our system and analyzing the performance of several real-world IoT applications on top of it.

\section{Related Work}
In this section, we briefly review a subset of relevant work to our system.

\fakeparagraph{Data Privacy \& Access Control.}
The current dominating approach for sharing data in the web services is based on the OAuth protocol~\cite{oauth} where a central trusted entity enforces user-defined access policies (does not fullfil \textit{R1}).
Sieve~\cite{sieve} addresses these shortcomings with a combination of key-homomorphic and attribute-based encryption schemes.
Many applications employ anonymized data collection~\cite{sweeney2002k} as an attempt to protect personally identifiable information.
However, researchers have demonstrated effective de-anonymization techniques~\cite{narayanan2008robust} which work even with a small set of high dimensionality data.

\fakeparagraph{Blockchain.}
In recent years, a new class of blockchain technologies have emerged
that utilize the accountable computing and auditability of blockchains for other domains.
Blockstack~\cite{Blockstack} introduces the concept of virtualchains and proposes a decentralized server-less DNS.
Blockstack extends to a decentralized public key distribution system and registry for user identities.
Storj~\cite{Storj} and FileCoin~\cite{Filecoin} {(secure successor of IPFS~\cite{ipfs})} introduce a distributed object storage. 
They are both targeted for archiving files and lack sharing features.
Enigma~\cite{Enigmafull, Enigma} is the closest to our approach in that it uses the blockchain for access control and enables sharing 
of off-chain stored data.
However, Enigma stores data access logs within the blockchain, without addressing the consequential scalability issues.
Moreover, their system does not accommodate for IoT stream data (not satisfing \textit{R3}).
Our approach is inspired by the above approaches, however, our focus on IoT data leads to a number of important design differences.

\fakeparagraph{IoT Storage.}
A few of our design decisions regarding IoT data streams are inspired by Bolt~\cite{Bolt}.
Bolt presents chunking of IoT data for performance gain and protects confidentiality of chunks.
However, Bolt relies on the cloud-centric model (does not address \textit{R1}).
``The Cloud is Not Enough"~\cite{GDP} discusses the pitfalls of the cloud-centric IoT and advocates a data-centric approach.
They leave concrete system proposals for future work.

\section{Conclusion}
In this paper,
we introduce the primary design of a distributed secure data storage system targeted for the Internet of Things.
Our system allows for fine-grained access control and sharing of time-series sensor data of various IoT applications.
Initial performance evaluation results are promising and show a moderate overhead due to our system.
We are currently in the process of finalizing a complete reference implementation of our system and building several IoT applications on top of it.

\balance
{\footnotesize \bibliographystyle{acm}
\bibliography{sigproc}

\end{document}